\documentclass[aps,prc,floatfix,twocolumn,groupedaddress,showpacs,showkeys]{revtex4}

\usepackage{epsfig}
\usepackage{graphicx}
\usepackage{amssymb}

 \newcommand{\beq}{\begin{equation}}
 \newcommand{\eeq}{\end{equation}}
 \newcommand{\beqa}{\begin{eqnarray}}
 \newcommand{\eeqa}{\end{eqnarray}}

\begin{document}


\title{$\phi$-meson production in  In-In collisions at $E_{\rm lab}$=158$A$ GeV: evidence for relics of a 
thermal phase.}

\author{E. Santini, H. Petersen, M. Bleicher}
\affiliation{Institut f\"ur Theoretische Physik, Goethe-Universit\"at,\\ 
 Max-von-Laue-Str.~1, D-60438 Frankfurt am Main, Germany\\}


\date{\today}

\begin{abstract}

Yields and transverse mass distributions of the $\phi$-mesons
reconstructed in the $\phi\rightarrow\mu^+\mu^-$ channel 
in In+In collisions at  $E_{\rm lab}$=158$A$ GeV are calculated within 
an integrated Boltzmann+hydrodynamics 
hybrid approach based on the Ultrarelativistic 
Quantum Molecular Dynamics (UrQMD) 
transport model with an intermediate hydrodynamic stage.
The analysis is performed for various centralities and a comparison with the 
corresponding NA60 data in the muon channel is presented.
We find that the hybrid model, that embeds an intermediate locally
equilibrated phase subsequently mapped into the transport 
dynamics according to thermal phase-space distributions, gives a good description of the 
experimental data, both in yield and slope. 
On the contrary, the pure transport model calculations tend to fail 
in catching the general properties of the $\phi$ meson production: 
not only the yield, but also the slope of the $m_T$ spectra, very poorly 
compare with the experimental observations. 
\end{abstract}


\maketitle

\section{Introduction}

The production of $\phi$ mesons is considered to be one of the key observables to 
probe the state of matter produced in relativistic heavy-ion collisions. 
Strangeness enhancement in relativistic nucleus-nucleus
collisions compared to nucleon-nucleon collisions
has been originally suggested as a possible signal for the formation of
a deconfined plasma of quarks and gluons during the initial 
state of the reaction 
\cite{Rafelski:1982pu,Koch:1986ud,Shor:1984ui}. The dominant production
of $s\bar{s}$ pairs via gluon-gluon interaction in the plasma
may result in an enhanced number of strange and multistrange
particles produced after hadronization; in particular, free
$s\bar{s}$ pairs would coalesce to form $\phi$ mesons 
\cite{Shor:1984ui}, whereas their
production in $pp$ collisions is suppressed according to the
Okubo-Zweig-Iizuka rule \cite{Okubo:1963fa,Zweig:1964,Iizuka:1966fk}. 
It is moreover expected that $\phi$ mesons decouple from the rest of the 
system earlier than other non-strange hadrons. 
At RHIC energies, an early decoupling of the $\phi$ 
from the hadronic rescattering dynamics was found in Refs. \cite{Cheng:2003as} and \cite{Hirano:2007ei}. 
Similar conclusions were obtained with the RQMD cascade in \cite{vanHecke:1998yu} for the 
$\Omega$ baryons at SPS energies.

The dominant hadronic decay of the $\phi$ meson is 
$\phi\rightarrow K \bar{K}$; 
additionally, being a neutral vector meson, the $\phi$ 
contributes to dilepton production via the direct decays  
$\phi\rightarrow e^+e^-$ and $\phi\rightarrow \mu^+\mu^-$.
Phi meson production was investigated extensively at 
the SPS by several experiments in 
the kaon (NA49 \cite{Afanasev:2000uu,Friese:2002re,Alt:2004wc,Alt:2008iv} 
and CERES \cite{Adamova:2005jr}), 
dielectron (CERES \cite{Adamova:2005jr}), and dimuon 
(NA50 \cite{Alessandro:2003gy}and NA60 \cite{Arnaldi:2009wr}) channels. 
The reconstruction of in-matter $\phi\rightarrow K \bar{K}$ decays, 
however, might be partially prevented by kaon absorption and 
rescattering \cite{Johnson:1999fv,Santini:2006cm,Filip:2001st} and a priori a careful investigation 
of kaon final state interactions cannot be avoided in a
quantitative comparison to experimental data.
On the contrary, the dilepton channel is 
not  affected by such a shortcoming: dileptons 
leave their production site essentially undistorted, and 
a comparison of model calculations to measurements in the dilepton channel 
is surely more straightforward. Recently, 
high statistics measurements of 
$\phi\rightarrow \mu^+\mu^-$ meson production in In-In collisions 
at  $E_{\rm lab}$=158$A$ GeV  have been  presented by the NA60 Collaboration 
\cite{Arnaldi:2009wr}. This reaction will be addressed in the present work.

To investigate  $\phi\rightarrow \mu^+\mu^-$ production 
in  In-In collisions at  $E_{\rm lab}$=158$A$ GeV we employ an integrated 
Boltzmann+hydrodynamics hybrid approach based on the Ultrarelativistic 
Quantum Molecular Dynamics (UrQMD) transport model with an intermediate 
hydrodynamic stage. In this approach, initial conditions and 
continuous decoupling up to freeze-out are treated within a 
full non-equilibrium transport approach, whereas 
hydrodynamics is used to describe the intermediate equilibrated phase.
This allows to reduce the parameters for the initial conditions and provides 
a consistent freeze-out description. 
Moreover, by comparing the hybrid approach to pure transport calculations, 
we are able to directly investigate the consequences that 
a dynamical approach involving local thermal and chemical equilibrium 
and one based on full non-equilibrium dynamics have on $\phi$ 
meson production.

The paper is structured in the following way: 
In Sec. \ref{The model}, we briefly discuss the hybrid model and 
present the procedure used to evaluate the  $\phi\rightarrow \mu^+\mu^-$ 
emission and, consequently, 
reconstruct the $\phi$ meson. 
In Sec. \ref{Results} we present calculations for $\phi$-meson 
transverse mass spectra as a function 
of centrality. The various contributions to the spectra associated 
with different stages of the reaction dynamics are shown and a 
comparison to the NA60 data is presented.
Finally, a summary and conclusions are given in Sec. \ref{Conclusions}.


\section{The model \label{The model}}

\subsection{The hybrid approach}

To simulate the dynamics
of the In+In collisions we employ a transport approach with an
embedded three-dimensional ideal relativistic one fluid
evolution for the hot and dense stage of the reaction
based on the UrQMD
 model \cite{Petersen:2008dd}. The present hybrid approach has 
been extensively described 
in  Ref. \cite{Petersen:2008dd}. Here, we limit ourselves 
to briefly describe  its main features and refer the reader to Ref. 
\cite{Petersen:2008dd} for details.

UrQMD \cite{Bass:1998ca,Bleicher:1999xi,Petersen:2008kb} is a hadronic transport approach which simulates multiple 
interactions of ingoing
and newly produced particles, the excitation and fragmentation
of color strings and the formation and decay
of hadronic resonances.  
The coupling between the
UrQMD initial state and the hydrodynamical evolution
proceeds when the two Lorentz-contracted nuclei have
passed through each other. Here, the spectators continue to propagate 
in the cascade and all other hadrons
are mapped to the hydrodynamic grid. This treatment
is especially important for non-central collisions which
are also studied in the present work. Event-by-event fluctuations are 
directly taken into account via initial conditions
generated by the primary collisions and string fragmentations in the 
microscopic UrQMD model. This leads to
non-trivial velocity and energy density distributions for
the hydrodynamical initial conditions \cite{Steinheimer:2007iy,Petersen:2009vx}. 
Subsequently, a full (3+1) dimensional
ideal hydrodynamic evolution is performed using
the SHASTA algorithm \cite{Rischke:1995ir,Rischke:1995mt}. The hydrodynamic
evolution is gradually merged into the hadronic cascade: 
to mimic an iso-eigentime hypersurface, full
transverse slices, of thickness $\Delta z$ = 0.2fm, are transformed to particles whenever in all
cells of each individual slice the energy density  drops below
five times the ground state energy density. 
The employment of such gradual transition allows to 
obtain a rapidity independent transition temperature 
without artificial time dilatation effects \cite{Petersen:2009gu} and has 
been explored in detail in various recent works 
\cite{Li:2008qm,Petersen:2009mz,Petersen:1900zz,Petersen:2009gu} 
devoted to SPS conditions. 
When merging, the hydrodynamic fields are transformed to particle
degrees of freedom via the Cooper-Frye equation in the computational
frame. The created particles proceed in their evolution in
the hadronic cascade where rescatterings and
final decays occur until all interactions cease and
the system decouples.

An input for the hydrodynamical calculation is
the equation of state (EoS). In this work we employ
a hadron gas equation of state, describing a
non-interacting gas of free hadrons \cite{Zschiesche:2002zr}. Included here
are all reliably known hadrons with masses up to $\approx$ 2
GeV, which is equivalent to the active degrees of freedom
of the UrQMD model.

\subsection{Reconstruction of the phi meson in the dimuon channel }
The reconstruction of the $\phi$ meson from the dimuon channel requires the 
evaluation of the $\phi\rightarrow\mu^+\mu^-$ emission. This is calculated 
perturbatively in the evolution stage that precedes or follows the 
hydrodynamical phase, and from thermal rates in the latter. In the following, 
the terms 
pre-equilibrium/pre-hydro (post-equilibrium/post-hydro), will be used 
to indicate the stage preceding (following) the mapping 
UrQMD$\rightarrow$hydrodynamical
 (hydrodynamical$\rightarrow$UrQMD) evolution description. 
Note that in the pre- and post-hydro stages the particles are the explicit 
degree of freedom and their interactions are explicitly treated within the cascade 
transport approach.  

\subsubsection{Reconstruction from pre-equilibrium and post-equilibrium emission}

Given the number $N^\phi_{\mu^+\mu^-}$ of $\mu^+\mu^-$ pairs emitted in  
$\phi\rightarrow \mu^+\mu^-$ decays, the number $N^{\rm rec}_{\phi}$ of 
$\phi$ meson reconstructed 
in the dimuon channel is 
\beq
N^{\rm rec}_{\phi}=N^\phi_{\mu^+\mu^-}/BR^\phi_{\mu^+\mu^-},
\label{eq1}
\eeq
where $BR^\phi_{\mu^+\mu^-}$ is the $\phi\rightarrow \mu^+\mu^-$ branching 
ratio. The latter is given by the ratio between the dimuon and the total width 
of the $\phi$ meson, i.e. $BR^\phi_{\mu^+\mu^-}=\Gamma^\phi_{\mu^+\mu^-}/
\Gamma^\phi_{\rm tot}$.
Thus,
\beq
N^{\rm rec}_{\phi}=N^\phi_{\mu^+\mu^-}~\frac{\Gamma^\phi_{\rm tot}}
{\Gamma^\phi_{\mu^+\mu^-}}.
\label{eq2}
\eeq
 
In the pre-equilibrium and post-equilibrium phase dimuon emission from the 
$\phi$-meson can be calculated perturbatively as
\beq
N^\phi_{\mu^+\mu^-}=\sum_n^{N_{\phi}}\int_{\tau_i(n)}^{\tau_f(n)}
\Gamma^\phi_{\mu^+\mu^-}~d\tau
\label{eq3}
\eeq
where now $N_{\phi}$ indicates the number of $\phi$ mesons present, 
in some stage of the evolution, in the system and  $\tau_i(n)$ ($\tau_f(n)$) 
are the times at which the 
$n$-th $\phi$ meson appeared in (disappeared from) the system and are 
evaluated in the meson rest-frame.
This perturbative method is known as time integration method or 
``shining method'' and has long been 
applied in the transport description of dilepton emission 
(see e.g. \cite{Li:1994cj,Vogel:2007yu,Schmidt:2008hm}).
Combining Eqs.(\ref{eq1})--(\ref{eq3}) one has:
\beq
N^{\rm rec}_{\phi}=\sum_n^{N_{\phi}}\int_{\tau_i(n)}^{\tau_f(n)}
\Gamma^\phi_{\rm tot}~d\tau
\label{cascmaster}
\eeq
which expresses the fact that the number of reconstructed mesons is 
proportional to the typical life-time of the meson in the system.

Some considerations are now in order. During the pre-equilibrium phase, 
$\tau_i(n)$ coincides with the time at which a $\phi$ meson is typically 
produced from a nucleon-nucleon scattering. With exception of some 
very rare almost instantaneous interaction, $\tau_f(n)$ typically coincides 
with the time at which the hydrodynamical phase starts. In other words, 
the probability of dimuon emission in the pre-hydro stage is evaluated up to 
the moment $\phi$ mesons are merged into 
the hydrodynamical phase.
From then on, $\phi$ emission will be treated as thermal. 

In the post-equilibrium phase, $\tau_i(n)$  coincides either with the time at 
which the 
transition  from the hydrodynamical to the transport description is 
performed (for those $\phi$ mesons produced 
via the Cooper-Frye equation) or with the time at which a $\phi$ meson is 
produced from hadronic interactions still occurring in the cascade phase, 
as result of the fact that the whole system is not yet completely decoupled; 
$\tau_f(n)$ is, in this case, the time at which the 
$\phi$ meson decays or eventually rescatters.

\subsubsection{Reconstruction from equilibrated thermal emission}
Thermal dimuon emission from the $\phi$ meson can be expected to be 
significantly smaller than the post-equilibrium emission, 
due to the fact that the lifetime of the fireball (7-10 fm) is much smaller 
than the  $\phi$ (vacuum) lifetime of $\sim$44 fm. 
To determine the thermal dimuon production rate from  $\phi$-meson decays 
we observe that the mass of the $\phi$ meson ($m_\phi$=1.019 GeV) is larger 
than the typical 
local temperature of the thermalized fireball, thus  
the particle number distribution function can be reasonably evaluated in 
Boltzmann approximation.
Moreover, the $\phi$ meson being a very narrow resonance 
($\Gamma^\phi_{\rm tot}=0.00426$ GeV and $\Gamma^\phi_{\rm tot}/m_\phi\approx 0.4\%$ ), for simplicity we 
neglect its small width and approximate the mass distribution of 
the meson with a $\delta(m^2-m_0^2)$ function (pole approximation).

In the Boltzmann and pole approximation the $\phi$ particle phase space 
distribution function is given by 
\beq
\frac{dN_\phi}{d^3x~d^3q} \left(q_0;T(\mathbf{x},t)\right)=\frac{3}{(2\pi)^3}~
e^{-q_0/T(\mathbf{x},t)},
\eeq
where $T(\mathbf{x},t)$ is the  local temperature, $(q_0,\mathbf{q})$ the meson 4-momentum in the thermal frame and the dependence of the 
temperature from the 
(discretized) space-time point of the (3+1) grid has 
been explicitly indicated. 
The number of dimuons produced per unit phase space volume and unit time from $\phi\rightarrow \mu^+\mu^-$ decays is
\beq
\frac{dN_{\mu^+\mu^-}}{d^4x~d^3q} \left(q_0;T(\mathbf{x},t)\right)=
\frac{dN_\phi}{d^3x~d^3q}~ \frac{\Gamma^\phi_{\mu^+\mu^-}}{\gamma_\phi(q_0)}.
\label{dimuonrate}
\eeq
Here $\Gamma^\phi_{\mu^+\mu^-}$ is the dimuon width as defined in the meson 
rest-frame and $\gamma_\phi=q_0/m_\phi$ is the Lorentz factor 
transforming from the meson to the thermal rest frame.
The number of dimuons emitted per unit space-time volume can be obtained by integrating Eq. (\ref{dimuonrate}) over momentum. The integration can be performed 
analytically and one finds: 
\beq
\frac{dN_{\mu^+\mu^-}}{d^4x} \left(T(\mathbf{x},t)\right)= \frac{3}{2\pi^2}~m_\phi^2~T(\mathbf{x},t)~
K_1\left( \frac{m_\phi}{T(\mathbf{x},t)}\right)\Gamma^\phi_{\mu^+\mu^-},
\label{momintrate}
\eeq 
where $K_1$ is the modified Bessel function of first order. 
The contribution of a single cell to the dimuon emission in the 
time step $\Delta t$ is therefore:
\beq
\left( N_{\mu^+\mu^-} \right)_{\rm cell}=\frac{dN_{\mu^+\mu^-}}{d^4x}\left(T_{\rm cell}\right)  ~V_{\rm cell} ~\Delta t.
\eeq
The cell contribution to the dimuon emission 
is calculated according to the above expression. 
To compare with experimental data, dimuon momenta in the c.m. frame are 
then generated with a Monte Carlo procedure according to the distribution function 
\beq
\frac{dN_{\mu^+\mu^-}}{d^3x~d^3p}\sim\frac{m_\phi}{p_0}~
f_B\left(\frac{p_\nu u^\nu}{T_{\rm cell}}\right)
\eeq
with  $u^\nu$ the fluid cell 4-velocity and $f_B$ the boson distribution function.

The total dimuon yield from thermal $\phi$ mesons is  
obtained summing the above expression over all fluid cells and all time 
steps of the (3+1) grid that are spanned by the system during the 
hydrodynamical evolution until the transition criterium is reached. 
Let us express this summation symbolically as:
\beq
N_{\mu^+\mu^-}^{\phi}=\int_{V\subset \mathbb{R}^4} d^4x~~
\left\{ \frac{3}{2\pi^2}~m_\phi^2~T(\mathbf{x},t)~
K_1\left( \frac{m_\phi}{T(\mathbf{x},t)}\right)\Gamma^\phi_{\mu^+\mu^-}
\right\}.
\eeq
The corresponding number of $\phi$ mesons reconstructed in the dimuon channel 
is then: 
\beq
N^{\rm rec}_{\phi}=\int_{V\subset \mathbb{R}^4}d^4x~~
\left\{ \frac{3}{2\pi^2}~m_\phi^2~T(\mathbf{x},t)~
K_1\left( \frac{m_\phi}{T(\mathbf{x},t)}\right)\Gamma^\phi_{\rm tot}
\right\}.
\label{thermalmaster}
\eeq

Before proceeding, we would like to make two considerations. 
The first one concerns an implication 
of the EoS used  for the hydrodynamical evolution.
As previously stated, a hadronic EoS has been used in the present work. 
This implies that dimuon emission from the $\phi$ meson is here 
evaluated during 
the whole hydrodynamical phase. In general, however, if a phase transition 
from quark-gluon to hadronic matter occurs, 
the hadronic thermal rate would be only a fraction of the total thermal rate.
In this sense, the results presented in the next section on thermal dimuon 
emission from the $\phi$ meson should be regarded as an upper limit 
of the possible amount of dimuons indeed produced by $\phi$ mesons during the 
high temperature/high density stage of the heavy-ion collision.

The second consideration regards 
the validity of the pole approximation used.
This approximation is justified as long as the $\phi$ meson maintains 
its properties of narrow resonance.
In medium, the $\phi$ meson is expected  to broaden, 
as suggested by hadronic many body calculations. 
If the amount of broadening is significant or/and the spectral 
function of the meson develops a complex ``structured'' shape in the medium, 
the pole approximation may loose validity. 
On a quantitative base, there is no general consensus 
on the specific amount of  broadening of the  $\phi$
meson spectral function to be expected in a high temperature/high 
density environment.  
Early evaluations of medium effects based on 
hadronic rescattering at finite temperature have indicated quite moderate 
changes of both the $\phi$ meson  mass and width \cite{Ko:1993id,Haglin:1994ap,
Haglin:1994xu,Smith:1997xu}. In a subsequent calculation, 
collision rates in a meson gas have been estimated to amount to a 
broadening by $\sim$  20 MeV at $T$=150 MeV \cite{AlvarezRuso:2002ib}. 
The dressing of the kaon cloud is presumably
the main effect for $\phi$ modifications in nuclear
matter, increasing its width by $\sim$ 25 MeV at normal nuclear
density \cite{Cabrera:2002hc}. In hot and baryon poor hadronic matter the 
in-medium  properties of the $\phi$ have been schematically explored in 
Ref. \cite{Rapp:2000pe}: the meson was found to retain its resonance structure 
and an in-medium width of $\sim$ 32 MeV at $(T,\mu_N)$=(180,27) MeV was estimated.
In a recent work \cite{Vujanovic:2009wr}, the spectral density of the 
$\phi$ meson in a hot bath of nucleons and pions has been 
microscopically calculated 
from the forward scattering amplitude in a two component approach. 
The authors found a considerable broadening of the meson width, e.g., 
$\Gamma^\phi_{\rm med}\sim$ 100 MeV at saturation density and temperature 
$T=150$ MeV.

In fact, there is no visible evidence for a strong in-medium scenario for 
the $\phi$ meson from the NA60 data. 
For all the analyzed centrality bins, the  measured  invariant 
mass distribution can be described in terms of a vacuum-like spectral function and the 
extracted values for the mass and the width are compatible with the PDG values and independent of centrality. Of course, one should remind that 
the extraction of the in-medium modified component of the total dimuon emission 
from the experimental data is a non-trivial task, since this component 
would lie under the large unmodified peak produced from the decays occurring at
the freeze-out.
Certainly, the study of in-medium modifications of the $\phi$ meson 
properties is in itself an interesting research topic. However,  
in the present work will not address this issue and assume that 
also in-medium the resonance maintains its narrow width, so that 
the pole approximation is still valid. 
For the meson pole mass $m_\phi$, moreover, the vacuum value will be used. 
In any case, as we will show below, from our analysis it emerges 
that the amount of $\phi$'s from thermal emission is by far smaller than the abundance from the cascade 
part, 
so that an eventual in-medium modification of the thermal rate 
will most likely not alter the results on the total transverse mass spectra 
discussed in the next section.

\section{Results \label{Results}}
In this section we investigate the relative abundances of 
$\phi\rightarrow \mu^+\mu^-$ production in the various stages 
of the system evolution and present results 
for $\phi$ meson invariant transverse mass spectra as a function of the collision centrality. Calculations for $\phi\rightarrow \mu^+\mu^-$ production in 
In-In collisions at  $E_{\rm lab}$=158$A$ GeV have been performed for 5 centrality classes.
 In agreement with the treatment of the experimental data, 
each class  was identified by the range of charged particle multiplicity 
$dN_{ch}/d\eta$ in the pseudorapidity interval 3$\leq\eta\leq$4.
The relation between a specific range of $dN_{ch}/d\eta$ and the corresponding centrality bin was specified by the NA60 collaboration and can be found in 
Table 1 of Ref. \cite{Arnaldi:2009wr}. The correspondence between the 
5 ranges of $dN_{ch}/d\eta$ and 5 ranges of impact parameters was obtained 
by the analysis of the charged particles obtained within the 
UrQMD hybrid model as a function of the impact parameter selected in the 
Monte Carlo simulation.

 \subsection{Relative abundances in the various evolution stages and $m_T$ spectra}

First, let us investigate the relation between the amount of $\phi$ mesons 
reconstructed from $\phi\rightarrow \mu^+\mu^-$ and the stage of the evolution 
probed by the dimuon emission. We start by discussing separately and in detail results for the most central bin. Later, an analogous analysis is presented as function of centrality.

\begin{figure}
\includegraphics[width=.46\textwidth]{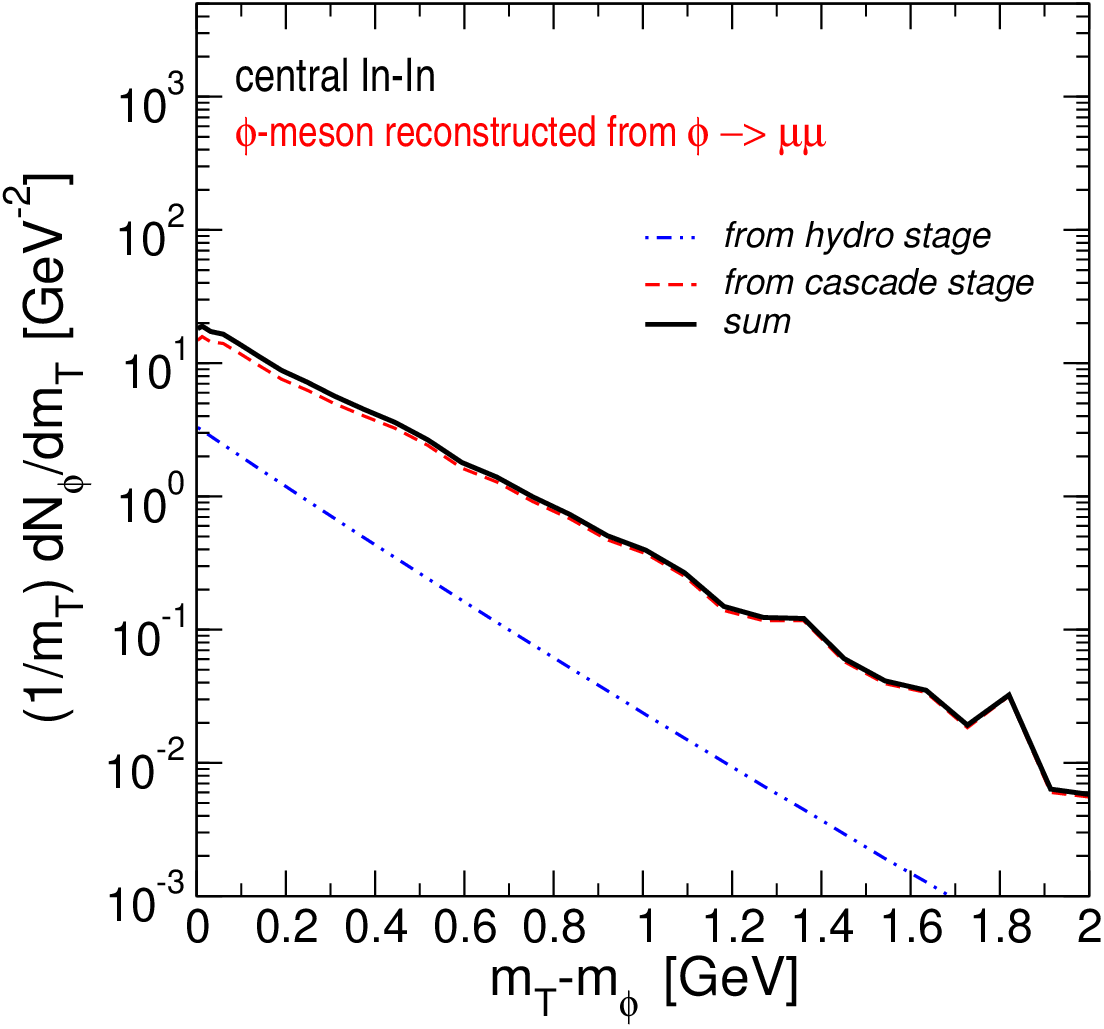}
\caption{Transverse mass distributions of the $\phi$ meson in central 
indium-indium collisions. The $\phi$ is reconstructed 
in the dimuon channel (see text).
Double-dotted-dashed line: contribution to the $\phi\rightarrow \mu^+\mu^- $ production of the hydrodynamical stage. Dashed line: contribution of the cascade stage. Full line: total   $\phi\rightarrow \mu^+\mu^- $ production. \label{fig1}}
\end{figure}

\begin{figure}
\includegraphics[width=.46\textwidth]{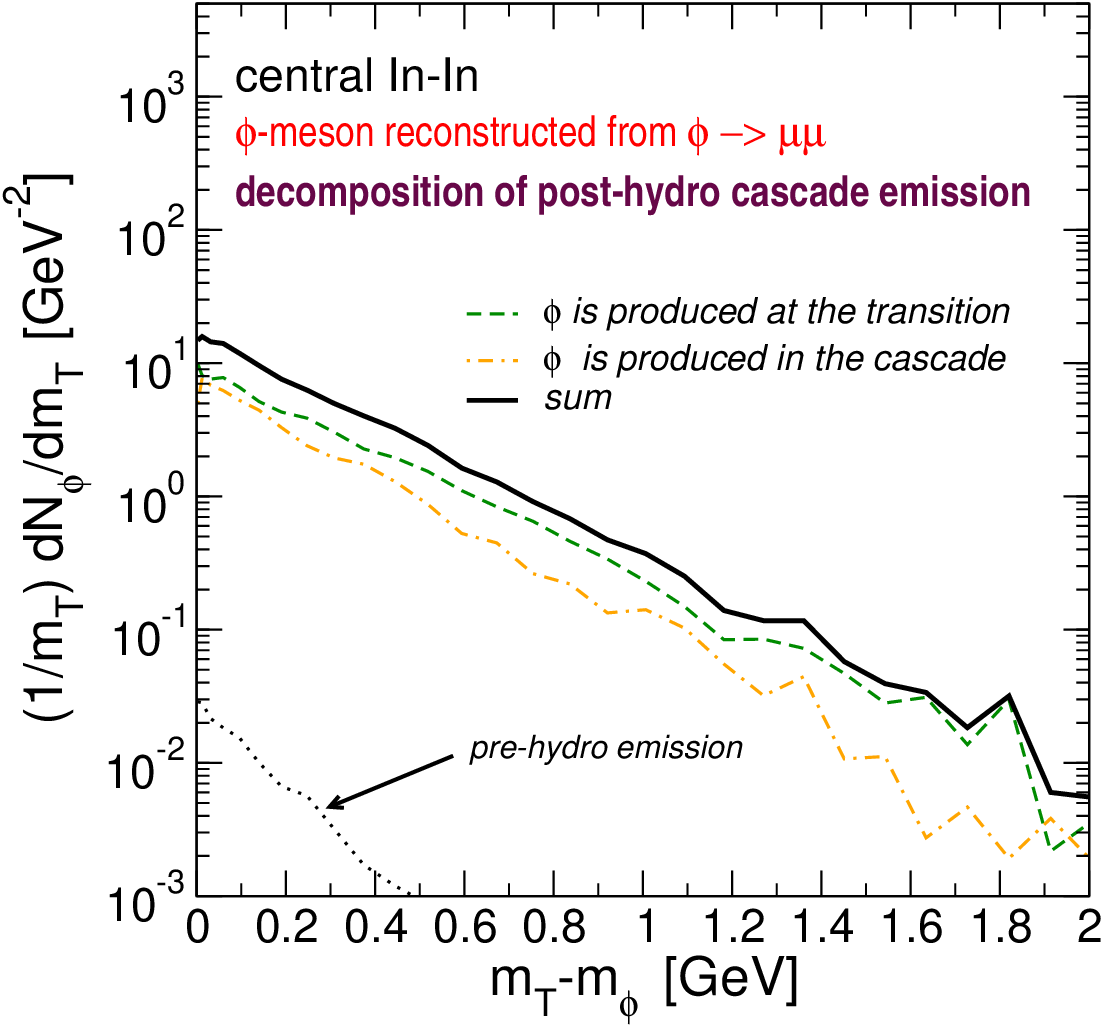}
\caption{ Decomposition of the post-equilibrium 
$\phi\rightarrow \mu^+\mu^- $ production in: (i) 
emission from  $\phi$ mesons which are merged into the cascade at the 
transition point via the Cooper-Frye equation and 
emit in the cascade stage of the evolution (dashed line); (ii) emission from 
$\phi$ mesons which are produced and emit in the cascade stage 
(dotted-dashed line). The pre-equilibrium emission is denoted by the dotted line.
The full line represents the total cascade emission. Due to the smallness of the 
pre-equilibrium emission the latter practically coincides with the 
total post-equilibrium emission. \label{fig1-b}}
\end{figure}

The contribution to the $\phi$ 
transverse mass spectra of 
those $\phi$ mesons emitting during the hydrodynamical and the 
cascade stage are separately shown in Fig.\ref{fig1}. 
The emission from the hydrodynamic stage is found almost one order of magnitude 
smaller than the emission from the cascade. As already mentioned, this is due 
to the relative smallness of the duration of the hydrodynamical phase when 
compared to the cascade phase. Thus, the dominant contribution comes from the 
cascade stage. 
This stage contains pre- and post-equilibrium emission. However, 
as shown in Fig.\ref{fig1-b}, the pre-equilibrium 
emission is negligible 
(two orders of magnitude smaller than 
the post-equilibrium emission) and will not be discussed further. 
The post-equilibrium emission can be divided in two categories: (i) the 
emission from $\phi$ particles produced via Cooper-Frye at the transition 
point and (ii) the emission from $\phi$ particles produced 
\emph{during} the cascade (Fig. \ref{fig1-b}). 
In the first case, the particles have a momentum distribution 
that reflects the thermal properties of the transition point, although 
their dimuon decay occurs later in a non-thermal environment. 
In this sense, this copious emission, though not specifically thermal 
(i.e. not described by thermal rate equations) still carries information about the preceding thermal phase. 
It represents the most direct ``remain'' of the thermally equilibrated 
phase previously experienced by the system.
The second contribution, on the contrary, can be labeled as a 
``purely cascade'' one. This is the contribution of $\phi$ particles produced in the 
non-equilibrium environment on the way to final decoupling. This second contribution is characterized by steeper transverse mass spectra. the shape of the total spectra is found to be composed by the interplay of both 
emissions.

\begin{figure}
 \includegraphics[width=.46\textwidth]{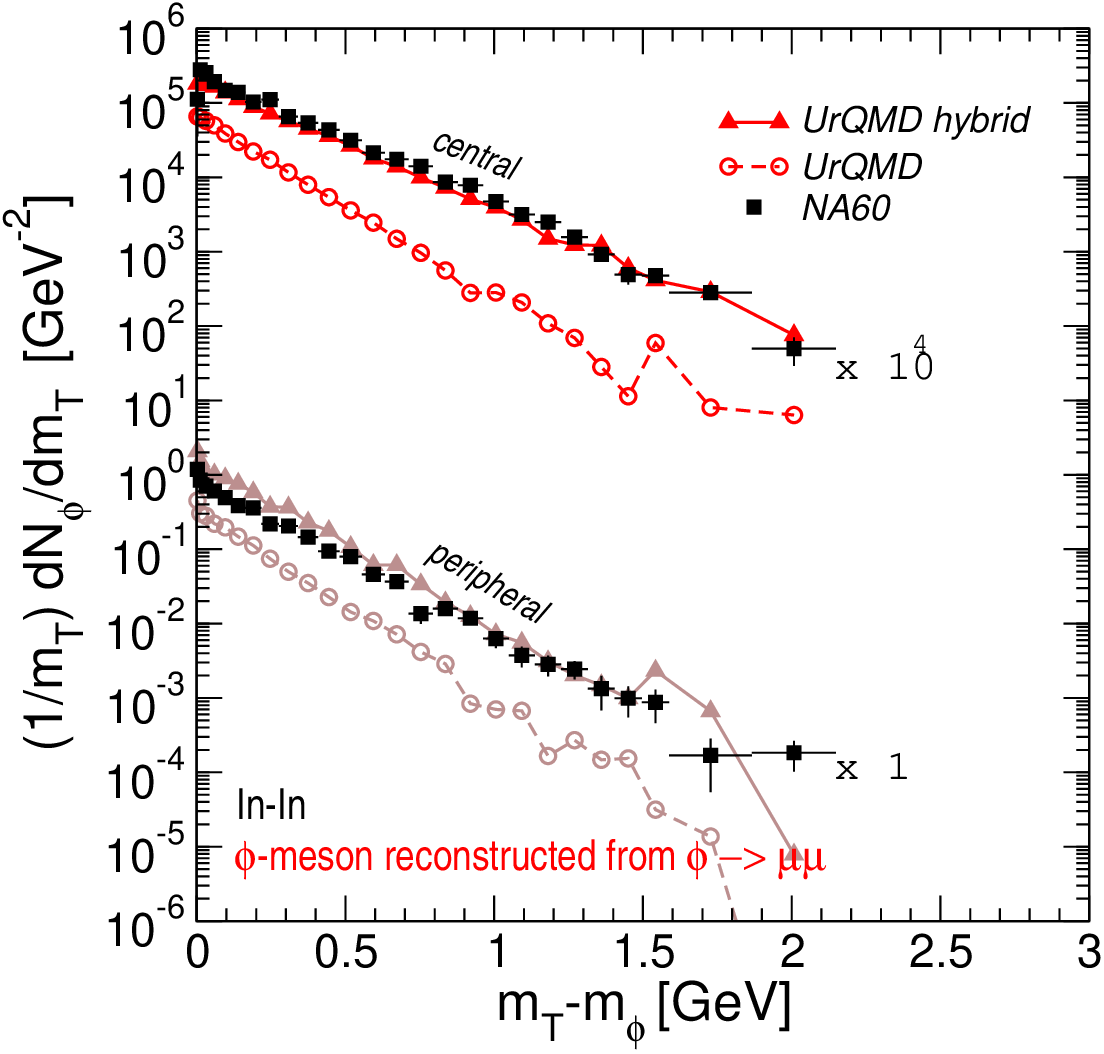}%
 \caption{Transverse mass distributions of the $\phi$ meson in central (top) and peripheral (bottom) indium-indium collisions. The hybrid model calculation (full line) is compared to the pure UrQMD transport calculation (dashed line) 
and to experimental data \cite{Arnaldi:2009wr}. Bin-widths which coincide with the ones of the experimental data have been here used. \label{compare}}
 \end{figure}

It is instructive to compare the hybrid model calculations to the pure 
cascade calculations, in which no assumption of an intermediate equilibrium 
phase is made. The comparison is presented in Fig. \ref{compare} (top) together with the experimental data. As one can see, 
the absence of an intermediate thermal phase results in a steepening of the 
transverse mass spectra not supported by the data. 
Moreover, the pure cascade calculation strongly underestimates the $\phi$ meson yield, a feature already emerged in recent independent investigations \cite{Alt:2008iv} performed in relation to measurements obtained by the NA49 experiment. 
There, it was found that a  statistical hadron gas model with undersaturation of strangeness \cite{Becattini:2005xt} could account for the measured yields \cite{Alt:2008iv}.

From this first analysis, we can conclude that, 
despite the smallness of the 
specifically thermal $\phi\rightarrow \mu^+\mu^-$ emission, the presence of the thermal phase is essential in order to obtain an appropriate yield 
and slope of the $m_T$ distributions from the cascade emission. In this sense, 
we speak about the presence of ``thermal relics''. 


\begin{figure}
 \includegraphics[width=.46\textwidth]{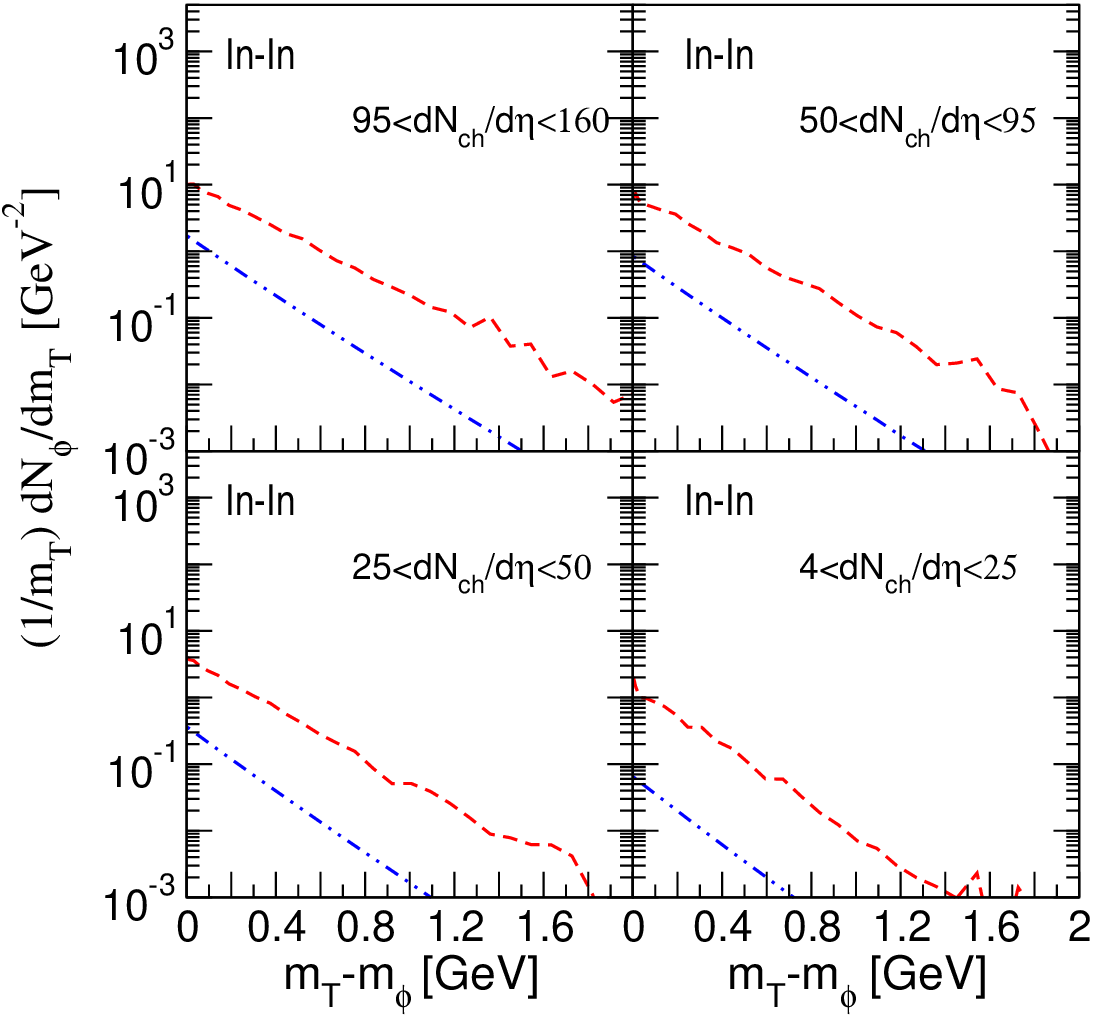}%
 \caption{Same as in Fig. \ref{fig1}, 
but for different centrality classes. Double-dotted-dashed line: contribution to the $\phi\rightarrow \mu^+\mu^- $ production of the hydrodynamical stage. Dashed line: contribution of the cascade stage. \label{fig3}}
 \end{figure}

\begin{figure}
 \includegraphics[width=.46\textwidth]{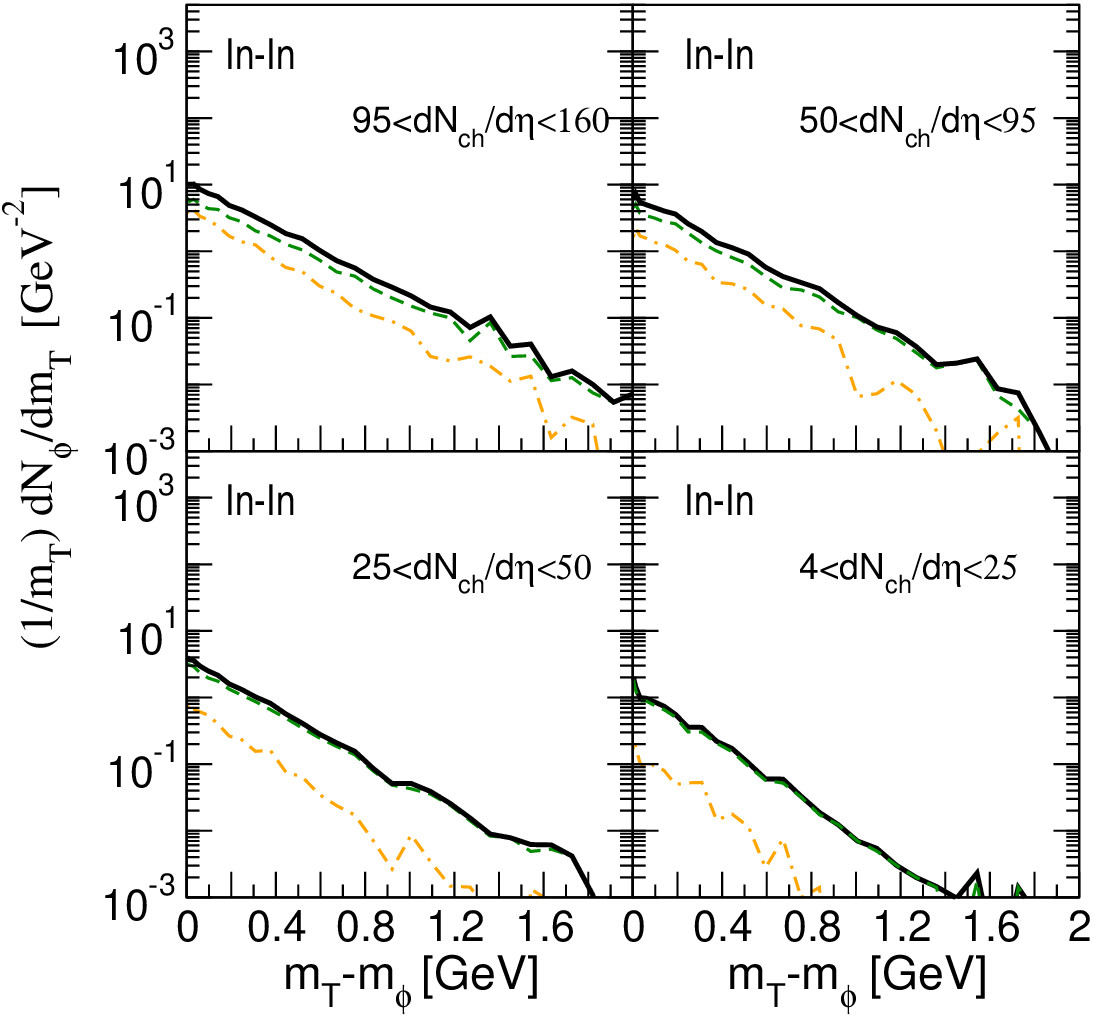}%
 \caption{Same as in Fig. \ref{fig1-b}, 
but for different centrality classes. Dashed line: emission from  $\phi$ mesons which are merged into the cascade at the 
transition point via the Cooper-Frye equation and 
emit in the cascade stage of the evolution; Dotted-dashed line: emission from 
$\phi$ mesons which are produced and emit in the cascade stage. \label{fig4}}
 \end{figure}

An analogous  analysis has been performed for the further 4 centrality bins 
and results are presented in Fig.\ref{fig3} and Fig.\ref{fig4}. 
For all centrality classes the thermal rate from the hydrodynamic evolution is found to be much smaller than the 
cascade emission. With decreasing centrality, the pure cascade 
emission (dash-dotted line) becomes less and less important and the spectra 
is more and more determined by the emission from those mesons emerging 
from the thermal stage into the cascade at the transition point.


 \begin{figure}
 \includegraphics[width=.46\textwidth]{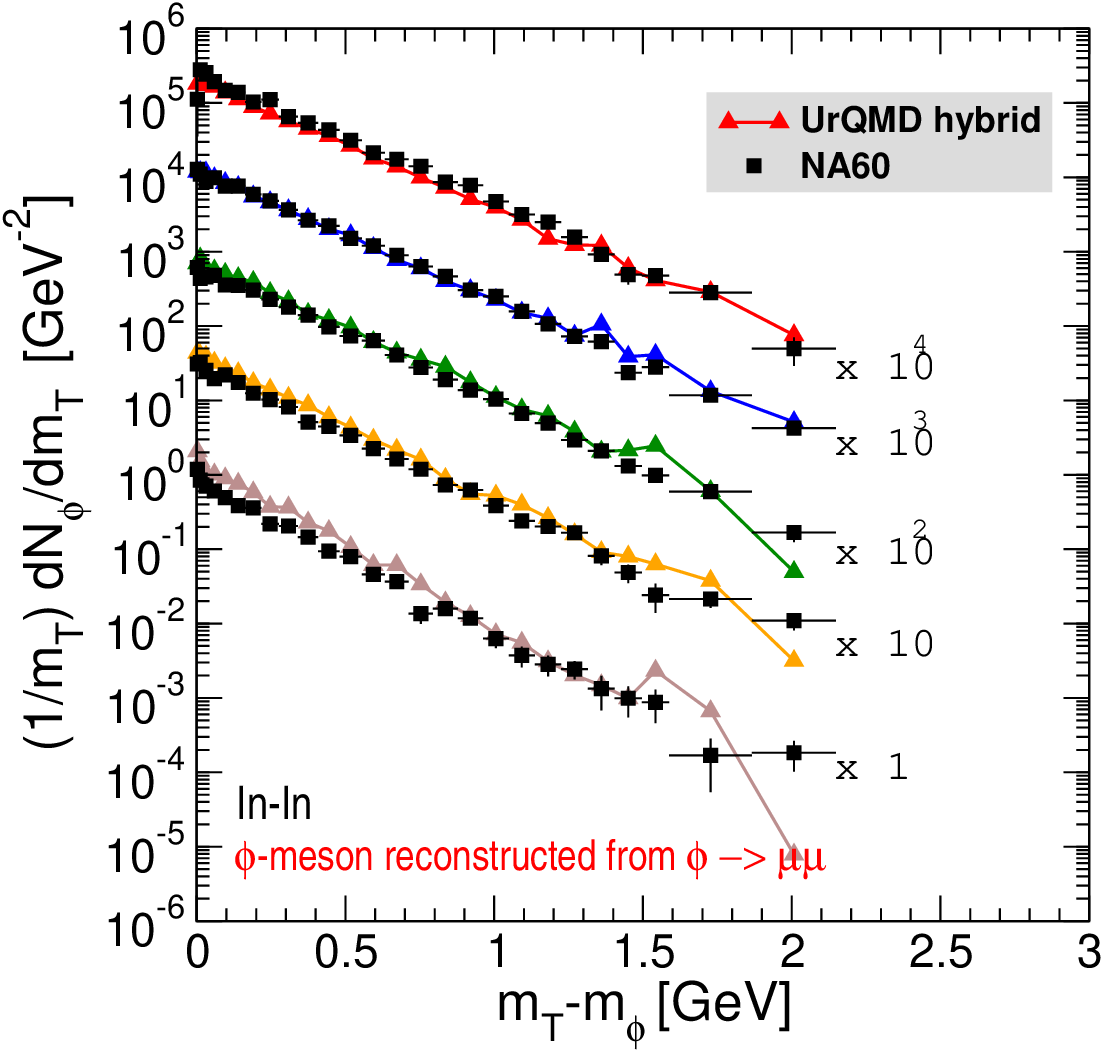}%
 \caption{Transverse mass distributions of the $\phi$ meson in indium-indium collisions as a function 
of centrality; from top to bottom: central to peripheral spectra. The hybrid model calculations are compared 
with NA60 data \cite{Arnaldi:2009wr}. Bin-widths which coincide with the ones of the experimental data have been here used. \label{mtspectra}}
 \end{figure}

Finally, the $\phi$ meson transverse mass spectra calculated within the 
hybrid model for the 5 centrality classes considered are compared to NA60 data in Fig. \ref{mtspectra}.  The hybrid model can account pretty well for both slope and yield in 
the  first 4 centrality classes. Small deviations  are observed for the utmost peripheral bin where, most likely, this kind of hybrid models are already at the limit of 
applicability. They rely on the assumption that an equilibrium phase is indeed reached, which for very peripheral reactions is at least questionable \cite{Petersen:2009zi}.

Pure transport calculations fail however too in describing this very peripheral 
reaction (see Fig. \ref{compare}, bottom), though the comparison with experimental data suggests that 
deviations of the resulting slope from the measured one are much smaller than 
the in the case of central collisions.

\section{\label{Conclusions}Summary and conclusions}
In this work we employed an integrated Boltzmann+hydrodynamics 
hybrid approach based on the Ultrarelativistic 
Quantum Molecular Dynamics  
transport model with an intermediate hydrodynamic stage 
to analyze $\phi$-meson production in In+In collisions at  $E_{\rm lab}$=158$A$ GeV from its 
reconstruction in the $\phi\rightarrow\mu^+\mu^-$ channel.
We find that the hybrid model fairly describes $\phi$ yields and 
transversal mass spectra at various 
collision centralities. 
In particular, the analysis points out that 
the underlying assumption of the existence of an 
intermediate equilibrated phase seems to play an essential role 
in order to  catch the main aspects of the physics emerging from 
the $\phi$ dimuon emission at top SPS energies. 

\begin{acknowledgments}
The authors acknowledge G. Torrieri for fruitful discussions and the 
NA60 Collaboration, M. Floris in particular, for providing the experimental data.
This work was supported by BMBF, GSI, DFG and the Hessen Initiative 
for Excellence (LOEWE) through the Helmholtz International Center for FAIR (HIC for FAIR). We thank the Center for Scientific Computing for providing computational resources.
\end{acknowledgments}

\bibliography{biblio}

\begin{thebibliography}{45}
\expandafter\ifx\csname natexlab\endcsname\relax\def\natexlab#1{#1}\fi
\expandafter\ifx\csname bibnamefont\endcsname\relax
  \def\bibnamefont#1{#1}\fi
\expandafter\ifx\csname bibfnamefont\endcsname\relax
  \def\bibfnamefont#1{#1}\fi
\expandafter\ifx\csname citenamefont\endcsname\relax
  \def\citenamefont#1{#1}\fi
\expandafter\ifx\csname url\endcsname\relax
  \def\url#1{\texttt{#1}}\fi
\expandafter\ifx\csname urlprefix\endcsname\relax\def\urlprefix{URL }\fi
\providecommand{\bibinfo}[2]{#2}
\providecommand{\eprint}[2][]{\url{#2}}

\bibitem[{\citenamefont{Rafelski and Muller}(1982)}]{Rafelski:1982pu}
\bibinfo{author}{\bibfnamefont{J.}~\bibnamefont{Rafelski}} \bibnamefont{and}
  \bibinfo{author}{\bibfnamefont{B.}~\bibnamefont{Muller}},
  \bibinfo{journal}{Phys. Rev. Lett.} \textbf{\bibinfo{volume}{48}},
  \bibinfo{pages}{1066} (\bibinfo{year}{1982}).

\bibitem[{\citenamefont{Koch et~al.}(1986)\citenamefont{Koch, Muller, and
  Rafelski}}]{Koch:1986ud}
\bibinfo{author}{\bibfnamefont{P.}~\bibnamefont{Koch}},
  \bibinfo{author}{\bibfnamefont{B.}~\bibnamefont{Muller}}, \bibnamefont{and}
  \bibinfo{author}{\bibfnamefont{J.}~\bibnamefont{Rafelski}},
  \bibinfo{journal}{Phys. Rept.} \textbf{\bibinfo{volume}{142}},
  \bibinfo{pages}{167} (\bibinfo{year}{1986}).

\bibitem[{\citenamefont{Shor}(1985)}]{Shor:1984ui}
\bibinfo{author}{\bibfnamefont{A.}~\bibnamefont{Shor}}, \bibinfo{journal}{Phys.
  Rev. Lett.} \textbf{\bibinfo{volume}{54}}, \bibinfo{pages}{1122}
  (\bibinfo{year}{1985}).

\bibitem[{\citenamefont{Okubo}(1963)}]{Okubo:1963fa}
\bibinfo{author}{\bibfnamefont{S.}~\bibnamefont{Okubo}},
  \bibinfo{journal}{Phys. Lett.} \textbf{\bibinfo{volume}{5}},
  \bibinfo{pages}{165} (\bibinfo{year}{1963}).

\bibitem[{\citenamefont{Zweig}(1964)}]{Zweig:1964}
\bibinfo{author}{\bibfnamefont{G.}~\bibnamefont{Zweig}}, \bibinfo{journal}{CERN
  report No.} \textbf{\bibinfo{volume}{8419/Th}}, \bibinfo{pages}{412}
  (\bibinfo{year}{1964}).

\bibitem[{\citenamefont{Iizuka}(1966)}]{Iizuka:1966fk}
\bibinfo{author}{\bibfnamefont{J.}~\bibnamefont{Iizuka}},
  \bibinfo{journal}{Prog. Theor. Phys. Suppl.} \textbf{\bibinfo{volume}{37}},
  \bibinfo{pages}{21} (\bibinfo{year}{1966}).

\bibitem[{\citenamefont{Cheng et~al.}(2003)\citenamefont{Cheng, Liu, Liu,
  Schweda, and Xu}}]{Cheng:2003as}
\bibinfo{author}{\bibfnamefont{Y.}~\bibnamefont{Cheng}},
  \bibinfo{author}{\bibfnamefont{F.}~\bibnamefont{Liu}},
  \bibinfo{author}{\bibfnamefont{Z.}~\bibnamefont{Liu}},
  \bibinfo{author}{\bibfnamefont{K.}~\bibnamefont{Schweda}}, \bibnamefont{and}
  \bibinfo{author}{\bibfnamefont{N.}~\bibnamefont{Xu}}, \bibinfo{journal}{Phys.
  Rev.} \textbf{\bibinfo{volume}{C68}}, \bibinfo{pages}{034910}
  (\bibinfo{year}{2003}).

\bibitem[{\citenamefont{Hirano et~al.}(2008)\citenamefont{Hirano, Heinz,
  Kharzeev, Lacey, and Nara}}]{Hirano:2007ei}
\bibinfo{author}{\bibfnamefont{T.}~\bibnamefont{Hirano}},
  \bibinfo{author}{\bibfnamefont{U.~W.} \bibnamefont{Heinz}},
  \bibinfo{author}{\bibfnamefont{D.}~\bibnamefont{Kharzeev}},
  \bibinfo{author}{\bibfnamefont{R.}~\bibnamefont{Lacey}}, \bibnamefont{and}
  \bibinfo{author}{\bibfnamefont{Y.}~\bibnamefont{Nara}},
  \bibinfo{journal}{Phys. Rev.} \textbf{\bibinfo{volume}{C77}},
  \bibinfo{pages}{044909} (\bibinfo{year}{2008}), \eprint{0710.5795}.

\bibitem[{\citenamefont{van Hecke et~al.}(1998)\citenamefont{van Hecke, Sorge,
  and Xu}}]{vanHecke:1998yu}
\bibinfo{author}{\bibfnamefont{H.}~\bibnamefont{van Hecke}},
  \bibinfo{author}{\bibfnamefont{H.}~\bibnamefont{Sorge}}, \bibnamefont{and}
  \bibinfo{author}{\bibfnamefont{N.}~\bibnamefont{Xu}}, \bibinfo{journal}{Phys.
  Rev. Lett.} \textbf{\bibinfo{volume}{81}}, \bibinfo{pages}{5764}
  (\bibinfo{year}{1998}), \eprint{nucl-th/9804035}.

\bibitem[{\citenamefont{Afanasev et~al.}(2000)}]{Afanasev:2000uu}
\bibinfo{author}{\bibfnamefont{S.~V.} \bibnamefont{Afanasev}}
  \bibnamefont{et~al.} (\bibinfo{collaboration}{NA49}), \bibinfo{journal}{Phys.
  Lett.} \textbf{\bibinfo{volume}{B491}}, \bibinfo{pages}{59}
  (\bibinfo{year}{2000}).

\bibitem[{\citenamefont{Friese}(2002)}]{Friese:2002re}
\bibinfo{author}{\bibfnamefont{V.}~\bibnamefont{Friese}}
  (\bibinfo{collaboration}{NA49}), \bibinfo{journal}{Nucl. Phys.}
  \textbf{\bibinfo{volume}{A698}}, \bibinfo{pages}{487} (\bibinfo{year}{2002}).

\bibitem[{\citenamefont{Alt et~al.}(2005)}]{Alt:2004wc}
\bibinfo{author}{\bibfnamefont{C.}~\bibnamefont{Alt}} \bibnamefont{et~al.}
  (\bibinfo{collaboration}{NA49}), \bibinfo{journal}{Phys. Rev. Lett.}
  \textbf{\bibinfo{volume}{94}}, \bibinfo{pages}{052301}
  (\bibinfo{year}{2005}), \eprint{nucl-ex/0406031}.

\bibitem[{\citenamefont{Alt et~al.}(2008)}]{Alt:2008iv}
\bibinfo{author}{\bibfnamefont{C.}~\bibnamefont{Alt}} \bibnamefont{et~al.}
  (\bibinfo{collaboration}{NA49}), \bibinfo{journal}{Phys. Rev.}
  \textbf{\bibinfo{volume}{C78}}, \bibinfo{pages}{044907}
  (\bibinfo{year}{2008}), \eprint{0806.1937}.

\bibitem[{\citenamefont{Adamova et~al.}(2006)}]{Adamova:2005jr}
\bibinfo{author}{\bibfnamefont{D.}~\bibnamefont{Adamova}} \bibnamefont{et~al.}
  (\bibinfo{collaboration}{CERES}), \bibinfo{journal}{Phys. Rev. Lett.}
  \textbf{\bibinfo{volume}{96}}, \bibinfo{pages}{152301}
  (\bibinfo{year}{2006}), \eprint{nucl-ex/0512007}.

\bibitem[{\citenamefont{Alessandro et~al.}(2003)}]{Alessandro:2003gy}
\bibinfo{author}{\bibfnamefont{B.}~\bibnamefont{Alessandro}}
  \bibnamefont{et~al.} (\bibinfo{collaboration}{NA50}), \bibinfo{journal}{Phys.
  Lett.} \textbf{\bibinfo{volume}{B555}}, \bibinfo{pages}{147}
  (\bibinfo{year}{2003}).

\bibitem[{\citenamefont{Arnaldi}(2009)}]{Arnaldi:2009wr}
\bibinfo{author}{\bibfnamefont{.~R.} \bibnamefont{Arnaldi}}
  (\bibinfo{collaboration}{The NA60}) (\bibinfo{year}{2009}),
  \eprint{0906.1102}.

\bibitem[{\citenamefont{Johnson et~al.}(2001)\citenamefont{Johnson, Jacak, and
  Drees}}]{Johnson:1999fv}
\bibinfo{author}{\bibfnamefont{S.~C.} \bibnamefont{Johnson}},
  \bibinfo{author}{\bibfnamefont{B.~V.} \bibnamefont{Jacak}}, \bibnamefont{and}
  \bibinfo{author}{\bibfnamefont{A.}~\bibnamefont{Drees}},
  \bibinfo{journal}{Eur. Phys. J.} \textbf{\bibinfo{volume}{C18}},
  \bibinfo{pages}{645} (\bibinfo{year}{2001}), \eprint{nucl-th/9909075}.

\bibitem[{\citenamefont{Santini et~al.}(2006)\citenamefont{Santini, Burau,
  Faessler, and Fuchs}}]{Santini:2006cm}
\bibinfo{author}{\bibfnamefont{E.}~\bibnamefont{Santini}},
  \bibinfo{author}{\bibfnamefont{G.}~\bibnamefont{Burau}},
  \bibinfo{author}{\bibfnamefont{A.}~\bibnamefont{Faessler}}, \bibnamefont{and}
  \bibinfo{author}{\bibfnamefont{C.}~\bibnamefont{Fuchs}},
  \bibinfo{journal}{Eur. Phys. J.} \textbf{\bibinfo{volume}{A28}},
  \bibinfo{pages}{187} (\bibinfo{year}{2006}), \eprint{nucl-th/0605041}.

\bibitem[{\citenamefont{Filip and Kolomeitsev}(2001)}]{Filip:2001st}
\bibinfo{author}{\bibfnamefont{P.}~\bibnamefont{Filip}} \bibnamefont{and}
  \bibinfo{author}{\bibfnamefont{E.~E.} \bibnamefont{Kolomeitsev}},
  \bibinfo{journal}{Phys. Rev.} \textbf{\bibinfo{volume}{C64}},
  \bibinfo{pages}{054905} (\bibinfo{year}{2001}), \eprint{hep-ph/0107288}.

\bibitem[{\citenamefont{Petersen
  et~al.}(2008{\natexlab{a}})\citenamefont{Petersen, Steinheimer, Burau,
  Bleicher, and Stocker}}]{Petersen:2008dd}
\bibinfo{author}{\bibfnamefont{H.}~\bibnamefont{Petersen}},
  \bibinfo{author}{\bibfnamefont{J.}~\bibnamefont{Steinheimer}},
  \bibinfo{author}{\bibfnamefont{G.}~\bibnamefont{Burau}},
  \bibinfo{author}{\bibfnamefont{M.}~\bibnamefont{Bleicher}}, \bibnamefont{and}
  \bibinfo{author}{\bibfnamefont{H.}~\bibnamefont{Stocker}},
  \bibinfo{journal}{Phys. Rev.} \textbf{\bibinfo{volume}{C78}},
  \bibinfo{pages}{044901} (\bibinfo{year}{2008}{\natexlab{a}}),
  \eprint{0806.1695}.

\bibitem[{\citenamefont{Bass et~al.}(1998)}]{Bass:1998ca}
\bibinfo{author}{\bibfnamefont{S.~A.} \bibnamefont{Bass}} \bibnamefont{et~al.},
  \bibinfo{journal}{Prog. Part. Nucl. Phys.} \textbf{\bibinfo{volume}{41}},
  \bibinfo{pages}{255} (\bibinfo{year}{1998}), \eprint{nucl-th/9803035}.

\bibitem[{\citenamefont{Bleicher et~al.}(1999)}]{Bleicher:1999xi}
\bibinfo{author}{\bibfnamefont{M.}~\bibnamefont{Bleicher}}
  \bibnamefont{et~al.}, \bibinfo{journal}{J. Phys.}
  \textbf{\bibinfo{volume}{G25}}, \bibinfo{pages}{1859} (\bibinfo{year}{1999}),
  \eprint{hep-ph/9909407}.

\bibitem[{\citenamefont{Petersen
  et~al.}(2008{\natexlab{b}})\citenamefont{Petersen, Bleicher, Bass, and
  Stocker}}]{Petersen:2008kb}
\bibinfo{author}{\bibfnamefont{H.}~\bibnamefont{Petersen}},
  \bibinfo{author}{\bibfnamefont{M.}~\bibnamefont{Bleicher}},
  \bibinfo{author}{\bibfnamefont{S.~A.} \bibnamefont{Bass}}, \bibnamefont{and}
  \bibinfo{author}{\bibfnamefont{H.}~\bibnamefont{Stocker}}
  (\bibinfo{year}{2008}{\natexlab{b}}), \eprint{0805.0567}.

\bibitem[{\citenamefont{Petersen and Bleicher}(2009)}]{Petersen:2009vx}
\bibinfo{author}{\bibfnamefont{H.}~\bibnamefont{Petersen}} \bibnamefont{and}
  \bibinfo{author}{\bibfnamefont{M.}~\bibnamefont{Bleicher}}
  (\bibinfo{year}{2009}), \eprint{0901.3821}.

\bibitem[{\citenamefont{Steinheimer et~al.}(2008)}]{Steinheimer:2007iy}
\bibinfo{author}{\bibfnamefont{J.}~\bibnamefont{Steinheimer}}
  \bibnamefont{et~al.}, \bibinfo{journal}{Phys. Rev.}
  \textbf{\bibinfo{volume}{C77}}, \bibinfo{pages}{034901}
  (\bibinfo{year}{2008}), \eprint{0710.0332}.

\bibitem[{\citenamefont{Rischke
  et~al.}(1995{\natexlab{a}})\citenamefont{Rischke, Bernard, and
  Maruhn}}]{Rischke:1995ir}
\bibinfo{author}{\bibfnamefont{D.~H.} \bibnamefont{Rischke}},
  \bibinfo{author}{\bibfnamefont{S.}~\bibnamefont{Bernard}}, \bibnamefont{and}
  \bibinfo{author}{\bibfnamefont{J.~A.} \bibnamefont{Maruhn}},
  \bibinfo{journal}{Nucl. Phys.} \textbf{\bibinfo{volume}{A595}},
  \bibinfo{pages}{346} (\bibinfo{year}{1995}{\natexlab{a}}),
  \eprint{nucl-th/9504018}.

\bibitem[{\citenamefont{Rischke
  et~al.}(1995{\natexlab{b}})\citenamefont{Rischke, Pursun, and
  Maruhn}}]{Rischke:1995mt}
\bibinfo{author}{\bibfnamefont{D.~H.} \bibnamefont{Rischke}},
  \bibinfo{author}{\bibfnamefont{Y.}~\bibnamefont{Pursun}}, \bibnamefont{and}
  \bibinfo{author}{\bibfnamefont{J.~A.} \bibnamefont{Maruhn}},
  \bibinfo{journal}{Nucl. Phys.} \textbf{\bibinfo{volume}{A595}},
  \bibinfo{pages}{383} (\bibinfo{year}{1995}{\natexlab{b}}),
  \eprint{nucl-th/9504021}.

\bibitem[{\citenamefont{Petersen
  et~al.}(2009{\natexlab{a}})\citenamefont{Petersen, Steinheimer, Burau, and
  Bleicher}}]{Petersen:2009gu}
\bibinfo{author}{\bibfnamefont{H.}~\bibnamefont{Petersen}},
  \bibinfo{author}{\bibfnamefont{J.}~\bibnamefont{Steinheimer}},
  \bibinfo{author}{\bibfnamefont{G.}~\bibnamefont{Burau}}, \bibnamefont{and}
  \bibinfo{author}{\bibfnamefont{M.}~\bibnamefont{Bleicher}}
  (\bibinfo{year}{2009}{\natexlab{a}}), \eprint{0907.2169}.

\bibitem[{\citenamefont{Petersen
  et~al.}(2009{\natexlab{b}})\citenamefont{Petersen, Steinheimer, Bleicher, and
  Stocker}}]{Petersen:2009mz}
\bibinfo{author}{\bibfnamefont{H.}~\bibnamefont{Petersen}},
  \bibinfo{author}{\bibfnamefont{J.}~\bibnamefont{Steinheimer}},
  \bibinfo{author}{\bibfnamefont{M.}~\bibnamefont{Bleicher}}, \bibnamefont{and}
  \bibinfo{author}{\bibfnamefont{H.}~\bibnamefont{Stocker}},
  \bibinfo{journal}{J. Phys.} \textbf{\bibinfo{volume}{G36}},
  \bibinfo{pages}{055104} (\bibinfo{year}{2009}{\natexlab{b}}),
  \eprint{0902.4866}.

\bibitem[{\citenamefont{Li et~al.}(2009)\citenamefont{Li, Steinheimer,
  Petersen, Bleicher, and Stocker}}]{Li:2008qm}
\bibinfo{author}{\bibfnamefont{Q.-f.} \bibnamefont{Li}},
  \bibinfo{author}{\bibfnamefont{J.}~\bibnamefont{Steinheimer}},
  \bibinfo{author}{\bibfnamefont{H.}~\bibnamefont{Petersen}},
  \bibinfo{author}{\bibfnamefont{M.}~\bibnamefont{Bleicher}}, \bibnamefont{and}
  \bibinfo{author}{\bibfnamefont{H.}~\bibnamefont{Stocker}},
  \bibinfo{journal}{Phys. Lett.} \textbf{\bibinfo{volume}{B674}},
  \bibinfo{pages}{111} (\bibinfo{year}{2009}), \eprint{0812.0375}.

\bibitem[{\citenamefont{Petersen
  et~al.}(2009{\natexlab{c}})\citenamefont{Petersen, Steinheimer, Burau, and
  Bleicher}}]{Petersen:1900zz}
\bibinfo{author}{\bibfnamefont{H.}~\bibnamefont{Petersen}},
  \bibinfo{author}{\bibfnamefont{J.}~\bibnamefont{Steinheimer}},
  \bibinfo{author}{\bibfnamefont{G.}~\bibnamefont{Burau}}, \bibnamefont{and}
  \bibinfo{author}{\bibfnamefont{M.}~\bibnamefont{Bleicher}},
  \bibinfo{journal}{Eur. Phys. J.} \textbf{\bibinfo{volume}{C62}},
  \bibinfo{pages}{31} (\bibinfo{year}{2009}{\natexlab{c}}).

\bibitem[{\citenamefont{Zschiesche et~al.}(2002)\citenamefont{Zschiesche,
  Schramm, Schaffner-Bielich, Stoecker, and Greiner}}]{Zschiesche:2002zr}
\bibinfo{author}{\bibfnamefont{D.}~\bibnamefont{Zschiesche}},
  \bibinfo{author}{\bibfnamefont{S.}~\bibnamefont{Schramm}},
  \bibinfo{author}{\bibfnamefont{J.}~\bibnamefont{Schaffner-Bielich}},
  \bibinfo{author}{\bibfnamefont{H.}~\bibnamefont{Stoecker}}, \bibnamefont{and}
  \bibinfo{author}{\bibfnamefont{W.}~\bibnamefont{Greiner}},
  \bibinfo{journal}{Phys. Lett.} \textbf{\bibinfo{volume}{B547}},
  \bibinfo{pages}{7} (\bibinfo{year}{2002}), \eprint{nucl-th/0209022}.

\bibitem[{\citenamefont{Li and Ko}(1995)}]{Li:1994cj}
\bibinfo{author}{\bibfnamefont{G.-Q.} \bibnamefont{Li}} \bibnamefont{and}
  \bibinfo{author}{\bibfnamefont{C.~M.} \bibnamefont{Ko}},
  \bibinfo{journal}{Nucl. Phys.} \textbf{\bibinfo{volume}{A582}},
  \bibinfo{pages}{731} (\bibinfo{year}{1995}), \eprint{nucl-th/9407016}.

\bibitem[{\citenamefont{Vogel et~al.}(2008)}]{Vogel:2007yu}
\bibinfo{author}{\bibfnamefont{S.}~\bibnamefont{Vogel}} \bibnamefont{et~al.},
  \bibinfo{journal}{Phys. Rev.} \textbf{\bibinfo{volume}{C78}},
  \bibinfo{pages}{044909} (\bibinfo{year}{2008}), \eprint{0710.4463}.

\bibitem[{\citenamefont{Schmidt et~al.}(2009)}]{Schmidt:2008hm}
\bibinfo{author}{\bibfnamefont{K.}~\bibnamefont{Schmidt}} \bibnamefont{et~al.},
  \bibinfo{journal}{Phys. Rev.} \textbf{\bibinfo{volume}{C79}},
  \bibinfo{pages}{064908} (\bibinfo{year}{2009}), \eprint{0811.4073}.

\bibitem[{\citenamefont{Ko and Seibert}(1994)}]{Ko:1993id}
\bibinfo{author}{\bibfnamefont{C.~M.} \bibnamefont{Ko}} \bibnamefont{and}
  \bibinfo{author}{\bibfnamefont{D.}~\bibnamefont{Seibert}},
  \bibinfo{journal}{Phys. Rev.} \textbf{\bibinfo{volume}{C49}},
  \bibinfo{pages}{2198} (\bibinfo{year}{1994}), \eprint{nucl-th/9312010}.

\bibitem[{\citenamefont{Haglin and Gale}(1994)}]{Haglin:1994ap}
\bibinfo{author}{\bibfnamefont{K.~L.} \bibnamefont{Haglin}} \bibnamefont{and}
  \bibinfo{author}{\bibfnamefont{C.}~\bibnamefont{Gale}},
  \bibinfo{journal}{Nucl. Phys.} \textbf{\bibinfo{volume}{B421}},
  \bibinfo{pages}{613} (\bibinfo{year}{1994}), \eprint{nucl-th/9401003}.

\bibitem[{\citenamefont{Haglin}(1995)}]{Haglin:1994xu}
\bibinfo{author}{\bibfnamefont{K.}~\bibnamefont{Haglin}},
  \bibinfo{journal}{Nucl. Phys.} \textbf{\bibinfo{volume}{A584}},
  \bibinfo{pages}{719} (\bibinfo{year}{1995}), \eprint{nucl-th/9410028}.

\bibitem[{\citenamefont{Smith and Haglin}(1998)}]{Smith:1997xu}
\bibinfo{author}{\bibfnamefont{W.}~\bibnamefont{Smith}} \bibnamefont{and}
  \bibinfo{author}{\bibfnamefont{K.~L.} \bibnamefont{Haglin}},
  \bibinfo{journal}{Phys. Rev.} \textbf{\bibinfo{volume}{C57}},
  \bibinfo{pages}{1449} (\bibinfo{year}{1998}), \eprint{nucl-th/9710026}.

\bibitem[{\citenamefont{Alvarez-Ruso and Koch}(2002)}]{AlvarezRuso:2002ib}
\bibinfo{author}{\bibfnamefont{L.}~\bibnamefont{Alvarez-Ruso}}
  \bibnamefont{and} \bibinfo{author}{\bibfnamefont{V.}~\bibnamefont{Koch}},
  \bibinfo{journal}{Phys. Rev.} \textbf{\bibinfo{volume}{C65}},
  \bibinfo{pages}{054901} (\bibinfo{year}{2002}), \eprint{nucl-th/0201011}.

\bibitem[{\citenamefont{Cabrera and Vicente~Vacas}(2003)}]{Cabrera:2002hc}
\bibinfo{author}{\bibfnamefont{D.}~\bibnamefont{Cabrera}} \bibnamefont{and}
  \bibinfo{author}{\bibfnamefont{M.~J.} \bibnamefont{Vicente~Vacas}},
  \bibinfo{journal}{Phys. Rev.} \textbf{\bibinfo{volume}{C67}},
  \bibinfo{pages}{045203} (\bibinfo{year}{2003}), \eprint{nucl-th/0205075}.

\bibitem[{\citenamefont{Rapp}(2001)}]{Rapp:2000pe}
\bibinfo{author}{\bibfnamefont{R.}~\bibnamefont{Rapp}}, \bibinfo{journal}{Phys.
  Rev.} \textbf{\bibinfo{volume}{C63}}, \bibinfo{pages}{054907}
  (\bibinfo{year}{2001}), \eprint{hep-ph/0010101}.

\bibitem[{\citenamefont{Vujanovic et~al.}(2009)\citenamefont{Vujanovic,
  Ruppert, and Gale}}]{Vujanovic:2009wr}
\bibinfo{author}{\bibfnamefont{G.}~\bibnamefont{Vujanovic}},
  \bibinfo{author}{\bibfnamefont{J.}~\bibnamefont{Ruppert}}, \bibnamefont{and}
  \bibinfo{author}{\bibfnamefont{C.}~\bibnamefont{Gale}}
  (\bibinfo{year}{2009}), \eprint{0907.5385}.

\bibitem[{\citenamefont{Becattini et~al.}(2006)\citenamefont{Becattini,
  Manninen, and Gazdzicki}}]{Becattini:2005xt}
\bibinfo{author}{\bibfnamefont{F.}~\bibnamefont{Becattini}},
  \bibinfo{author}{\bibfnamefont{J.}~\bibnamefont{Manninen}}, \bibnamefont{and}
  \bibinfo{author}{\bibfnamefont{M.}~\bibnamefont{Gazdzicki}},
  \bibinfo{journal}{Phys. Rev.} \textbf{\bibinfo{volume}{C73}},
  \bibinfo{pages}{044905} (\bibinfo{year}{2006}), \eprint{hep-ph/0511092}.

\bibitem[{\citenamefont{Petersen
  et~al.}(2009{\natexlab{d}})\citenamefont{Petersen, Mitrovski, Schuster, and
  Bleicher}}]{Petersen:2009zi}
\bibinfo{author}{\bibfnamefont{H.}~\bibnamefont{Petersen}},
  \bibinfo{author}{\bibfnamefont{M.}~\bibnamefont{Mitrovski}},
  \bibinfo{author}{\bibfnamefont{T.}~\bibnamefont{Schuster}}, \bibnamefont{and}
  \bibinfo{author}{\bibfnamefont{M.}~\bibnamefont{Bleicher}}
  (\bibinfo{year}{2009}{\natexlab{d}}), \eprint{0903.0396}.

\end{thebibliography}

\end{document}